\documentclass[twocolumn,showpacs,preprintnumbers,amsmath,amssymb]{revtex4}

\usepackage{graphicx}
\usepackage{dcolumn}
\usepackage{bm}

\begin{document}

\preprint{Accepted by Phys. Rev. A}

\title{The Population Oscillation of Multicomponent \\
Spinor Bose-Einstein Condensate Induced by Nonadiabatic Transitions}

\author{Xiuquan Ma}
\author{Lin Xia}
\author{Fang Yang}
\author{Xiaoji Zhou}
\author{Yiqiu Wang}
\author{Hong Guo}\thanks{Author to whom correspondence should be addressed. E-mail:
hongguo@pku.edu.cn, phone: +86-10-6275-7035, Fax: +86-10-6275-3208. }
\author{Xuzong Chen}\thanks{Author to whom correspondence should be
addressed. E-mail: xuzongchen@pku.edu.cn, phone: +86-10-6275-1778, Fax: +86-10-6275-3208.} \affiliation{Key
Laboratory for Quantum Information and Measurements, Ministry of Education, School of Electronics Engineering and
Computer Science, Peking University, Beijing 100871, P. R. China\\}

\date{\today}

\begin{abstract}
The generation of the population oscillation of the multicomponent spinor Bose-Einstein condensate is demonstrated
in this paper. We observe and examine the nonsynchronous decreasing processes of the magnetic fields generated
between quadrupole coils and Ioffe coils during the switch-off of the quadrupole-Ioffe-configuration trap, which
is considered to induce a nonadiabatic transition. Starting from the two-level Schr\"{o}dinger equation, we have
done some numerical fitting and derived an analytical expression identical to the results of E. Majorana and C.
Zener, of which both the results well match the experimental data.
\end{abstract}

.

\maketitle

\section{introduction}
Recently, the multicomponent spinor Bose-Einstein condensate (BEC) has become a ``hot'' topic, since it shows an
appealing expectation in providing entangled spin systems applied in quantum optics and quantum computations
\cite{you2000,pu2000}, and in the quantized votex applied in the study of superconductors and superfluidity
\cite{cornell2004,machida2004}. Since more features \cite{ho2000,klausen2001,ueda2002,you2003,zhang2002} and more
phenomena \cite{ma2005,wadati2004,turkey2003,russia2003} are explored, the quantitatively manipulating and
splitting the multicomponent spinor BEC is urgently desired. In several groups, the means of optical manipulation
has been applied to split and study the spinor BEC \cite{stamper1998,stamper1999,chapman2004,schmaljohann2004}.
However, since the BEC is more commonly generated in static magnetic traps, the splitting and manipulating by the
means of magnetic field will be more convenient and efficient \cite{ma2005}.

Nonadiabatic transitions of atoms within magnetic sublevels are very old problems established in about 1930s
\cite{guettinger1931,laundau1932,zener1932}. The model of Majorana transition in which the magnetic field evolves
as $B_{x}(t)=0,B_{y}(t)={\rm Const},B_{z}(t)=kt$ and $t=(-\infty,\infty)$ was created and solved by E. Majorana in
1932 \cite{majorana1932}. Then, a more explicit and easily comprehensible explanation about the Majorana
transition was described by I. I. Rabi \cite{rabi1937}. After that, though further theoretical discussions and
analyses and other implements such as group theories have emerged to improve the study of Majorana transition
\cite{schwinger1937,rabi1939,bloch1940,rabi1945,salwen1955,formula1958}, few quantitative experiments nor
theoretical investigations have been undertaken, because there were not any precise ways to control the fleeting
hot atoms, and some qualitative explanations about spin flips and atom loss in magnetic quadrupole traps are well
known \cite{cornell1995,ketterle1995,spinflip1997}. However, with the development in the experiments and
techniques of ultracold atoms, the almost motionless atoms can be provided to interact with the swiftly rotating
magnetic field, and it makes possible the quantitative examination of Majorana transitions. In our previous work,
we demonstrated that Majorana transitions will emerge during the switch-off of the magnetic trap in our experiment
system and induce the split multicomponent spinor BEC \cite{ma2005}.

\begin{figure}\label{figver}
\centering
\includegraphics[width=9cm]{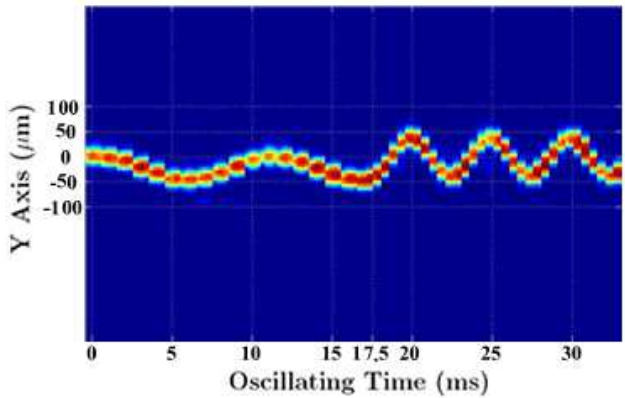}
\caption{(Color online) The Vertical Oscillation: The $y$-direction coordinate labels versus the vertical
oscillating time in the trap. The vertical down is the direction of the gravity. Different oscillating frequency
is due to different tightness of trapping confinement. The oscillation before the time $17.5$ ms represents the
switch-off of the compensate coils, while after it represents the restoring of the compensate coils. When the
compensate coils are switched off, the average velocity of the oscillation is around $1$ cm/s, and the period is
calculated to be around 11 ms.}
\end{figure}

In this paper, we report the observation and the explanation of population oscillation of the multicomponent
spinor BEC induced by nonadiabatic transitions. We believe that the formation of the population oscillation is
caused by the vertical oscillation of the condensate cloud and the nonsynchronous decreasing processes (NDP) of
the magnetic fields, which are both measured experimentally and demonstrated in this paper. Further, we derive an
analytical expression of this, starting from the Schr\"{o}dinger equation and the experimental conditions, while
the numerical calculation has also been undertaken. Both the analytical and numerical fitting results well fit the
experiment data.

\section{Generating the population oscillation of multicomponent spinor BEC}
Our experiment is set up on a standard equipment system for generating a cigar-like condensate in a dilute gas of
$^{87}$Rb. A compact low-power quadrupole-Ioffe-configuration (QUIC) trap with trapping frequencies of
$\omega_{r}=2\pi\times220$ Hz in radial directions and $\omega_{z}=2\pi\times20$ Hz in the axial direction, in
which the quadrupole coils are assumed to be along $x$-direction and the Ioffe coils along the $z$-direction, is
mounted on the lower chamber where the vacuum is up to $2\times10^{-11}$ mbar. After the evaporative cooling, the
condensate cloud is held in the center of magnetic trapping potential generated by $23.6$ A current in Ioffe coils
and $24.1$ A current in quadrupole coils. In addition, in order to reduce the offset field $B_{0}$ in the center
of the trap, we also apply a couple of compensate coils along the $z$-direction to regulate the tightness of the
trap center. When the compensate coils are switched on, the $B_{0}$ is reduced from $8.5$ Gauss to $1.5$ Gauss and
the typical gradient of our trap is about 150 Gauss/cm.

\begin{figure}
\centering\label{figexp}
\includegraphics[width=9cm]{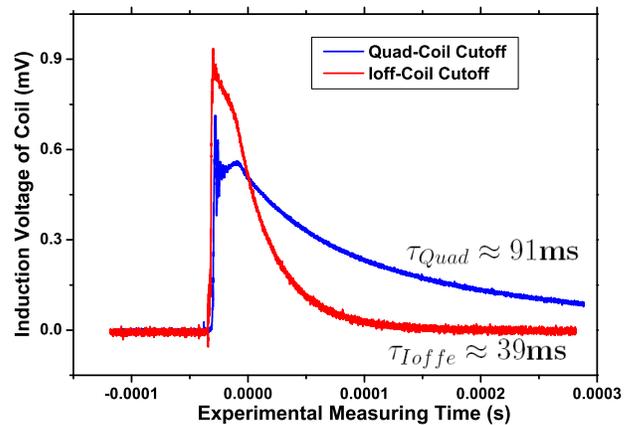}
\caption{(Color online) The Measurement of Nonsynchronous Decreasing Processes (NDP) : Induced voltage of the
detective coil when the quadrupole and Ioffe coils are shut down versus time during the decreasing process of the
currents. We applied a detective coil to examine the evolution of the field generated by the two types of coils
respectively. While one is measured, the other is substituted by an inductance vessel. The blue line represents
the detecting results of quadrople coils, while the red one represents that of Ioffe coils. The time constant of
exponential evolution is respectively about $91$ ms and $39$ ms.}
\end{figure}

Firstly, the compensate coils are switched off, and after a certain time interval, which we can call ``coil delay
time'', the QUIC trap (the quadrupole coils and Ioffe coils) is switched off afterwards. During this coil delay
time, as we know, the condensate cloud will be oscillating up and down vertically due to the balance of gravity
and new trapping potential whose confinement is much tighter than the original one due to the change of $B_{0}$.
To demonstrate the vertical oscillation clearly, we first switch off the compensate coils and then restore it (see
Fig.~1). We can see that, at different coil delay time during the oscillation, the condensate cloud may stay at
the different altitude which can be expressed
\begin{eqnarray}\label{ysite}
Y_{vertical}({\rm \mu m})=23.2\times\cos0.564 t ({\rm ms})-28.2,
\end{eqnarray}
where the positive direction of $y$-direction is vertically up (opposite to that of gravity) . From both the
expression Eq.~(\ref{ysite}) and the Fig.~1, we find that the average velocity of the oscillation is around 1 cm/s
and the period is calculated to be around 11 ms.

The fact is, though the quadrupole trap and Ioffe trap are switched off at the same the time, their decreasing
processes are totally nonsynchronous according to our experimental measurement (see Fig.~2). We apply a detective
coil to detect the field evolution of the two types of coils respectively. While one is measured by the detective
coil, the other is substituted by an inductance vessel. This means of detection can not give the time constant of
evolution of the real field but the changing trend of the field. Due to the derivative feature of the exponential
function, the time constant we detected can be seen to be identical to that of the real field.

As our previous work shows \cite{ma2005}, when the NDP of magnetic fields emerges, the condensate atoms will
experience the zero field $B_{z}=0$ and the reversion of the direction. In this case, there comes a nonadiabatic
transition among magnetic sublevels, and the different components of spinor BEC will be separated in space by
Stern-Gerlach Effect due to the interaction between atom spins and the gradient of magnetic field. The population
oscillation is shown partly from 18ms of coil delay time to 28ms of coil delay time (see Fig.~3), and surely much
longer oscillation has been observed in our experiment. Examining all the pictures during the whole process, we
find out that the period of the population oscillation is about $11$ ms. The total number of the atoms is about
$2\times10^{5}$. The most left cloud is considered to represent the $m_{F}=+2$ component and the most right one
belongs to $m_{F}=-2$ component, for we assume that the direction of the field after the reversion is the quantum
axis. It is necessary to point out that the periods of the vertical oscillations and population oscillation are
more or less the same. We believe this means that the population oscillation is induced by the vertical
oscillation and the different distributions of atoms are related to the different altitudes of the condensate
cloud.

\section{the basic model for two-level nonadiabatic transition}

\begin{figure}\label{figcolor}
\centering
\includegraphics[width=5.5cm]{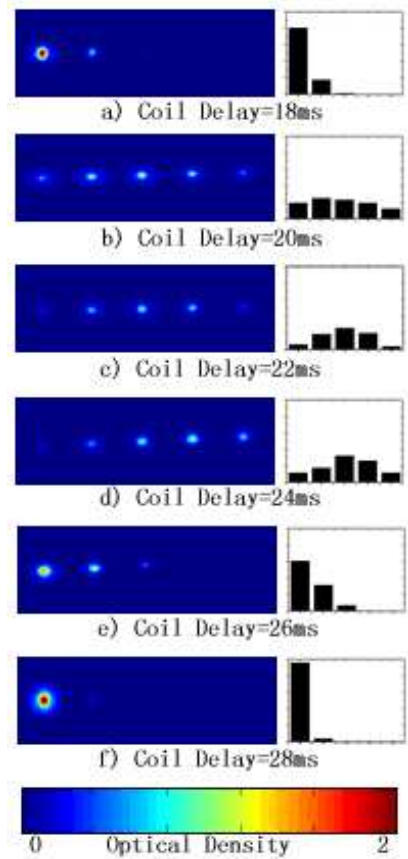}
\caption{(Color online) The Population Oscillation: Images of multicomponent spinor BEC due to different ``coil
delay times''. Each component of spinor BEC is separated by Stern-Galach effect due to the interaction between
atom spin and the gradient of the magnetic field. The bar charts of population distribution of each image are also
shown. From a) to f), the pictures are one period of the oscillation and the period value is examined to be around
$11$ms. The color bar shows the optical density of each pixel on the absorption images. We consider the most left
cloud representing the $m_{F}=+2$ component and the most right one representing the $m_{F}=-2$ component, for we
assume the direction of the field that is after the reversion to be the quantum axis. }
\end{figure}

According to I. I. Rabi's description, the Majorana transitions only take place in the case that the rotating
frequency of the magnetic field $f_{Rot.}=\partial B(t)/2\pi\partial t$ is big enough to be comparable to the
Larmor frequency of the field $f_{Lar.}=g\mu_{0}B(t)/2\pi\hbar$ \cite{rabi1937}. This is easy to understand from
the perspective of magnetic resonant transitions (MRT). Because the emergence of the zero field and the reversion
of the direction often bring the small magnitude of the field and huge rotating frequency, the nonadiabatic
transition will happen more easily. In our experiment, the evolution of the magnetic field just coincides with
this conclusion (see Fig.~4). When the field reverses its direction, the rotating frequency is remarkably huge. At
this time, the magnetic moment of the atom cannot follow the rotating field and there will be a transition among
the magnetic sublevels that can be ascribed to the Majorana transition.

Majorana Formula has been derived from both quantum mechanics and group theories to explain the multilevel cases
\cite{majorana1932,formula1958}. For a system with a total angular moment $J$,
\begin{eqnarray}\label{majo}
\nonumber&&P_{m,m'}=(J+m)!(J+m')!(J-m)!(J-m')!\left(\cos\frac{\theta}{2}\right)^{4J}\\
&&\times\left[\sum_{\nu=0}^{2J}\frac{(-1)^{\nu}(\tan\frac{\theta}{2})^{2\nu-m+m'}}{\nu!(\nu-m+m')!
(J-m-\nu)!(J-m'-\nu)!}\right]^{2},
\end{eqnarray}
where the value of the parameter $\theta$ is given by the two-level transition
\begin{eqnarray}\label{majo2}
\sin^{2}\frac{\theta}{2}=P_{\frac{1}{2},-\frac{1}{2}}.
\end{eqnarray}
This means that once we solve the two-level case, the results can be generalized for any system with any value of
$J$. So, in this part, we will focus on the physical model of the two-level nonadiabatic transition.

\begin{figure}
\centering
\includegraphics[width=9cm]{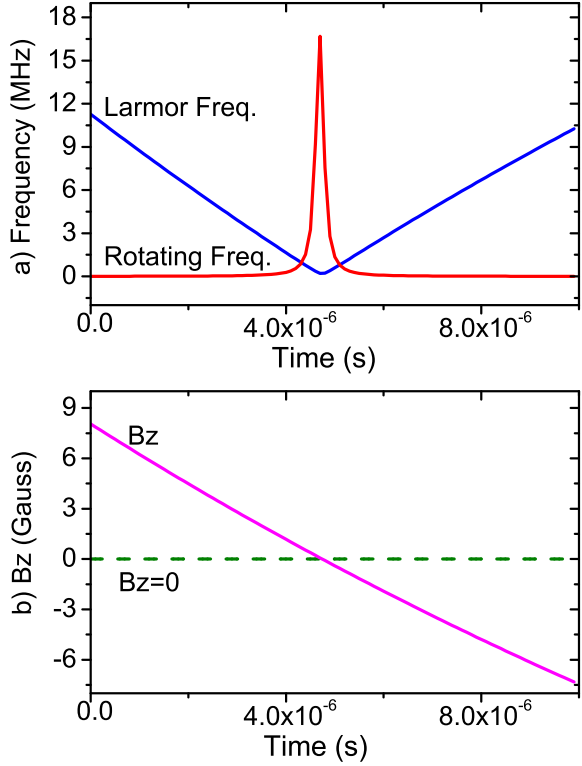}
\caption{(Color online) a) The Comparison of The Two Frequencies: The Larmor frequency $f_{Lar.}$ is proportional
to the magnitude of the field, so when the field reverse its direction, it is small. Meanwhile, the rotating
frequency $f_{Rot.}$ is remarkably huge due to the reversion. The time, during which $f_{Rot.}$ is big enough to
be comparable to the $f_{Lar.}$, is about 1 $\mu$s. b) The Real Evolution of The Magnetic Field $B_{z}$: When the
QUIC trap is switched off, the zero field $B_{z}=0$ emerges around 5 $\mu$s. Though the decreasing processes of
the field are exponential due to the discharge of the coils, the part at which the transition happens is
approximately linear. From these two figures, it is confirmed that the emergence of $B_{0}=0$ and the reversion of
the direction will induce the nonadiabatic Majorana transition.}
\end{figure}

The time, during which $f_{Rot.}$ is big enough to be comparable to the $f_{Lar.}$, is about 1 $\mu$s (see
Fig.~4a), so the movement of the cloud is about 0.01 $\mu$m according to the average velocity about 1 cm/s.
Compared with the vertical oscillation magnitude ($\sim$ 50 {\rm $\mu$m}), atom cloud can be seen to be
motionless. So we consider a system of motionless atoms whose spin moment is $s=\frac{1}{2}$ with a time evolving
magnetic field $\vec{B}(t)$. In this simple case, we begin with the Schr\"{o}dinger equation
\begin{equation}\label{basic}
i\hbar \left( \begin{array}{c}
  \dot{c_{1}} \\
  \dot{c_{2}} \\\end{array} \right)
=\frac{g\mu_{B}}{\hbar} \hat{\vec{F}}\cdot\vec{B}(t) \left(\begin{array}{c}
                                c_{1}\\
                                c_{2}\\\end{array}\right),
\end{equation}
where we know $\hat{\vec{F}}=\frac{\hbar}{2}\hat{\vec{\sigma}}$ in the two-level case. According to the
configuration of QUIC trap, the center of the magnetic trap is right on the $z$-axis. In addition to the dragging
down of the gravity, however, the center of the total trapping potential is slightly down along y-direction. So,
the three components of the evolving magnetic field should take the form $\vec{B}(t)=(0, B_{y}(t),B_{z}(t))$.
Putting the matrix form into the basic equation Eq.~(\ref{basic}) and taking the substitution $a=g\mu_{B}/2\hbar$,
one yields
\begin{eqnarray}\label{yeqs}
\left\{
\begin{array}{l}
  \dot{c_{1}}=-iaB_{z}c_{1}-aB_{y}c_{2}, \\
  \dot{c_{2}}=aB_{y}c_{1}+iaB_{z}c_{2},\\
\end{array}
\right.
\end{eqnarray}
and the initial conditions are:
\begin{eqnarray}\label{ychu}
\left\{
\begin{array}{l}
c_{1}(0)=1, \\
c_{2}(0)=0.
\end{array}
\right.
\end{eqnarray}
Substituting the real evolution of the magnetic field into Eq.~(\ref{yeqs}), one can get the transition
probability.

Though, as we know, the evolution of the magnetic field of the coils is exponentially down due to the discharging
process of the coils, i.e.,
\begin{eqnarray}
\left\{
\begin{array}{l}
B_{y}(t)=B_{yi}\exp\left(-\tau_{i} t\right)+B_{yq}\exp\left(-\tau_{q} t\right), \\
B_{z}(t)=B_{zi}\exp\left(-\tau_{i} t\right)-B_{zq}\exp\left(-\tau_{q} t\right),
\end{array}
\right.
\end{eqnarray}
where $B_{yi}$ and $B_{zi}$ are respectively the $y$-direction and $z$-direction component of magnetic field
generated from the Ioffe coils, $B_{yq}$ and $B_{zq}$ are from the quadrupole coils, and $\tau_{i}$ and $\tau_{q}$
are the reciprocals of the exponential time constants $\tau_{Ioffe}$ and $\tau_{Quad}$ in Fig.~2. Since only the
transition properties will be taken into account, it is appropriate to take the first order approximation
$\exp\left(-\tau_{q,i}t\right)\approx1-\tau_{q,i}t$ and describe them in the linear forms (see Fig.~4b)
\begin{eqnarray}\label{bsim1}
\left\{
\begin{array}{l}
B_{y}(t)\approx A_{y}-C_{y}t, \\
B_{z}(t)\approx A_{z}-C_{z}t,
\end{array}
\right.
\end{eqnarray}
where
\begin{eqnarray}
\left\{
\begin{array}{l}
A_{y}=\left(B_{yi}+B_{yq}\right),\\
A_{z}=\left(B_{zi}-B_{zq}\right),\\
C_{y}=\left(B_{yi}\tau_{i}+B_{yq}\tau_{q} \right),\\
C_{z}=\left(B_{zi}\tau_{i}-B_{zq}\tau_{q} \right).
\end{array}
\right.
\end{eqnarray}

From above we know $\tau_{i}\approx1/39$ms is twice of $\tau_{q}\approx1/91$ms (see Fig.~2) and hence
$\tau_{i}\approx 2\tau_{q}$. So the value of $C_{y}$ and $C_{z}$ can be estimated to be $C_{y}\approx
\tau_{q}\left(2B_{yi}+B_{yq} \right)$ and $C_{z}\approx \tau_{q}\left(2B_{zi}-B_{zq} \right)$, and the typical
values of theirs satisfy $C_{z}\gg C_{y}$. So we can simplify the expressions of magnetic fields by fixing the
$B_{y}$ at the value $A_{y0}=A_{y}-C_{y}\cdot t_{0}$ at the time $t=t_{0}$ when the field $B_{z}$ reverses its
direction at $B_{z}(t_{0})=0$,
\begin{eqnarray}\label{byz}
\left\{
\begin{array}{l}
B_{y}(t)\approx A_{y0}, \\
B_{z}(t)\approx A_{z}-C_{z}t.
\end{array}
\right.
\end{eqnarray}
With Eq.~(\ref{byz}) and all other substitutions taken into the Eq.~(\ref{yeqs}), we can get the second-order
differential equations which can be transformed into Webber equations \cite{zener1932}. As referred above, the
transition only occurs when $B_{z}$ reverses its direction, so it is allowed to reset the initial condition so as
to utilize the asymptotic solutions of Webber equations
\begin{eqnarray}\label{ini2}
\left\{
\begin{array}{l}
c_{1}(-\infty)=1,\\
c_{2}(-\infty)=0.
\end{array}
\right.
\end{eqnarray}
At the same time, the transition probability should take the form $P=1-|c_{2}(+\infty)|^{2}$, since the magnetic
field has reversed its direction \cite{majorana1932,zener1932}.

Therefore, we can derive the analytical expression of the Webber equation with its infinite asymptotic solutions.
If we take this substitution
\begin{eqnarray}\label{twore}
\alpha=\frac{aA_{y0}^{2}}{2C_{z}},
\end{eqnarray}
the final result of the analytical expression is (the calculation details can be found in \cite{zener1932} and
\cite{kleppner1981} ) :
\begin{eqnarray}
P_{\frac{1}{2},-\frac{1}{2}}=\exp(-2\pi \alpha).
\end{eqnarray}
The above analytic expression is identical to the results derived previously \cite{majorana1932,zener1932}. From
this analytical solution, the population oscillation should rely on the magnitude of $A_{y0}$ which is the only
unfixed parameter in $\alpha$. In our experiment, $A_{y0}$ is approximately proportional to altitude oscillation
$Y_{vertical}$, because the value of $B_{yi}+B_{yq}$ is around linear in radial directions. Hence, we
approximately have $\alpha\approx k Y^{2}_{vertical}$ and $k$ is a constant. Finally, we know that the transition
probability will also oscillate with the altitude oscillation $Y_{vertical}$
\begin{eqnarray}
P_{\frac{1}{2},-\frac{1}{2}}\approx\exp\left(-2\pi k Y^{2}_{vertical}\right),
\end{eqnarray}
which confirms that the population oscillation is induced by the vertical oscillation. As mentioned above, the
distribution of atom population among the split multicomponent spinor BEC can be obtained by the two-level result
Eq.~(\ref{twore}) and Majorana Formula Eq.~(\ref{majo}) and Eq.~(\ref{majo2}).

\begin{figure}\label{figre}
\centering
\includegraphics[width=9cm]{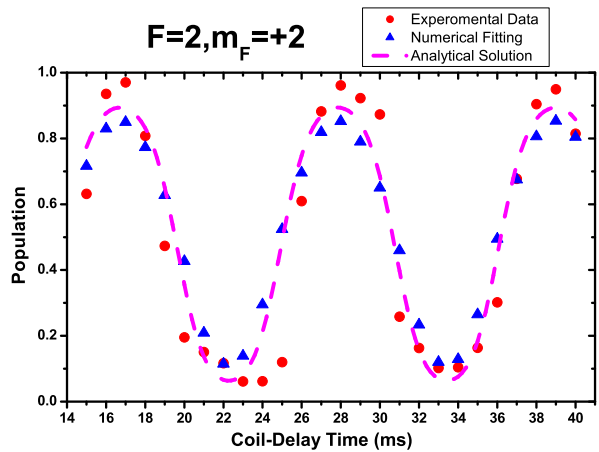}
\caption{(Color online) The Experimental Data and Analyzing Results: The population oscillation of
$|F=2,m_{F}=+2\rangle$ state versus coil delay time. The red circles represent the experimental data, while the
blue triangles represent the numerical fitting. The dashed line is drawn by the analytic solution from above
theoretical deducing. }
\end{figure}

The numerical fitting of the nonadiabatic transition has also been done by directly taking the exponential
expressions of magnetic field $\vec{B}(t)$ into the basic Schr\"{o}dinger equation Eq.~(\ref{basic}) or
Eq.~(\ref{yeqs}). To demonstrate the precision of our analytic analysis, the experiment data, numerical fitting
results and theoretical deducing results are illustrated at the same time (see Fig.~5). Though each sublevel state
has its own oscillation pattern, we only illustrate the $|F=2, m_{F}=+2\rangle$ state for instance. Since the
experimental points before $15$ms of coil delay time are slightly effected by the eddy current of coils and
mountings, we merely count the data ranging from 15ms to $40$ms.

\section{conclusion}
To conclude, we observed the split spinor BEC and the population oscillation of the five components of $^{87}$Rb
generated in the magnetic trap. For the first time, according to our knowledge, the population oscillation in
which the population distribution of the five components can be regulated by setting the experiment parameters is
reported and explained for its experimental formation. We believe that our experiment could most probably offer a
novel and profound means to produce and manipulate the spinor BEC and the distributions of its components. In
addition, we analyze the phenomenon of BEC splitting, which coincides with the explanation of NDP proposed by ANU
group \cite{close2002}, and measure the value of the DNP experimentally.

Starting with Schr\"{o}dinger equation, together with the experimental conditions, we derive the analytical
solution of nonadiabatic Majorana transition of condensate atoms within the magnetic sublevels. Though Majorana
transition has been broadly applied in cold atom physics for qualitative explanations, it is for the first time
that we quantitatively apply it to well fit our experimental data of the nonadiabatic transition.

For further experiments, we have designed a more complex system to control the magnetic fields of the two types of
coils separately, and with the aids of the theoretical analysis mentioned above, the arbitrary manipulations of
the atom populations among the multicomponent spinor BEC are anticipated. Besides, we also plan to load the
separated multicomponent condensates into the optical trap and study the interaction between the different
components. To summarize, our experimental method provides a convincing and potential way of studying the
ultracold atom physics, including BEC and atom optics.

\section*{Acknowlegements}
We thank Dr. Shuai Chen for his wonderful previous work on our experiment system and helpful discussions with us.
We thank Professor J\"{o}rg Schmiedmayer and Dr. Stephan Schneider for their helpful suggestion in our experiment.
We thank Mr. Lin Yi for his technical supports on LabVIEW programming work. This work is supported by the National
Fundamental Research Programme of China under Grant No. 2001CB309308, and No. 2005CB3724503, the Major Program of
National Natural Science Foundation of China under Grant No. 60490280, National Natural Science Foundation of
China under Grant No. 60271003 and 10474004, the National High Technology Research and Development Program of
China (863 Program) international cooperation program under Grant No. 2004AA1Z1220, and partially by the DAAD
Exchange Grogram (D/0212785 Personenaustausch VR China) and and DAAD exchange pro- gram: D/05/06972
Projektbezogener Personenaustausch mit China (Germany/China Joint Research Program).


\begin{references}
\bibitem{you2000} L. You and M. S. Chapman, Phys. Rev. A \textbf{62}, 052302 (2000).
\bibitem{pu2000} H. Pu and P. Meystre, Phys. Rev. Lett. \textbf{85}, 3987 (2000).
\bibitem{cornell2004} V. Schweikhard, I. Coddington, P. Engels, S. Tung, and and E. A. Cornell, Phys. Rev. Lett.
\textbf{93}, 210403 (2004).
\bibitem{machida2004} Takeshi Mizushima, Naoko Kobayashi, and Kazushige Machida, Phys. Rev. A \textbf{70}, 043613
(2004).
\bibitem{ho2000} Tin-Lun Ho and Lan Yin, Phys. Rev. Lett. \textbf{84}, 2302 (2000).
\bibitem{klausen2001} Nille N. Klausen, John L. Bohn and Chris H. Greene, Phys. Rev. A \textbf{64}, 053602 (2001).
\bibitem{ueda2002} Masahito Ueda and Masato Koashi, Phys. Rev. A \textbf{65}, 063602 (2002).
\bibitem{you2003} B. Zeng, D. L. Zhou, P. Zhang, Z. Xu, and L. You, Phys. Rev. A \textbf{68}, 042316 (2003).
\bibitem{zhang2002} Ping Zhang, C. K. Chan, Xiang-Gui Li, Qi-Kun Xue, and Xian-Geng Zhao, Phys. Rev. A
\textbf{66}, 043606 (2002).
\bibitem{ma2005} X. Q. Ma, S. Chen, F. Yang, L. Xia, X. J. Zhou, Y. Q. Wang, and X. Z. Chen, Chin. Phys. Lett.
\textbf{22}, 1106 (2005).
\bibitem{wadati2004} Jun'ichi Ieda, Takahiko Miyakawa, and Miki Wadati, Phys. Rev. Lett. \textbf{93}, 194102
(2002).
\bibitem{turkey2003} \"{O}. E. M\"{u}stecaplio\u{g}lu, M. Zhang, S. Yi, L. You, and C. P. Sun, Phys. Rev. A
\textbf{68}, 063616 (2003).
\bibitem{russia2003} Evgeny N. Bulgakov and Almas. F. Sadreev, Phys. Rev. Lett. \textbf{90}, 200401 (2003).
\bibitem{stamper1998} D. M. Stamper-Kurn, M. R. Andrews, A. P. Chikkatur, S. Inouye, H.-J. Miesner, J. Stenger,
and W. Ketterle, Phys. Rev. Lett. \textbf{80}, 2027 (1998).
\bibitem{stamper1999} D. M. Stamper-Kurn, H.-J. Miesner, A. P. Chikkatur, S. Inouye, J. Stenger,
and W. Ketterle, Phys. Rev. Lett. \textbf{83}, 661 (1999).
\bibitem{chapman2004} M.-S. Chang, C. D. Hamley, M. D. Barrett, J. A. Sauer, K.M. Fortier, W. Zhang, L. You,
and M. S. Chapman, Phys. Rev. Lett. \textbf{92}, 140403 (2004).
\bibitem{schmaljohann2004} H. Schmaljohann, M. Erhard, J. Kronj\"{a}ger, M. Kottke, S. van Staa, L. Cacciapuoti,
J. J. Arlt, K. Bongs, and K. Sengstock, Phys. Rev. Lett. \textbf{92}, 040402 (2004).
\bibitem{guettinger1931} P. G\"{u}ttinger, Zeits. f. Physik \textbf{73}, 169 (1931).
\bibitem{laundau1932} L. D. Laundau, Phys. Z. Sowjetunion \textbf{2}, 46 (1932); L. D. Landau and E. M. Lifshitz,
\emph{Quantum Mechanics: Nonrelativistic Theory}, 3rd ed. (Pergamon, New York, 1977), pp.342-351.
\bibitem{zener1932} C. Zener, Proc. Roy. Soc. London Ser. \textbf{A137}, 696 (1932).
\bibitem{majorana1932} E. Majorana, Nuovo Cimento \textbf{9}, 43 (1932).
\bibitem{rabi1937} I. I. Rabi, Phys. Rev. \textbf{51}, 652 (1937).
\bibitem{schwinger1937} J. Schwinger, Phys. Rev. \textbf{51}, 648 (1937).
\bibitem{rabi1939} I. I. Rabi, S. Millman, and P. Kusch, Phys. Rev. \textbf{55}, 526 (1939).
\bibitem{bloch1940} F. Bloch and A. Siegert, Phys. Rev. \textbf{57}, 522 (1940).
\bibitem{rabi1945} F. Bloch and I. I. Rabi, Rev. Mod. Phys. \textbf{17}, 237 (1945).
\bibitem{salwen1955} Harold. Salwen, Phys. Rev. \textbf{99}, 1274 (1955).
\bibitem{formula1958} Alvin. Meckler, Phys. Rev. \textbf{111}, 1447 (1958).
\bibitem{cornell1995} M. Anderson, J. Ensher, M. Matthews, C. Wieman, and E. Cornell, Science \textbf{269}, 198 (1995).
\bibitem{ketterle1995} K. Davis, M. Mewes, M. Andrews, N.J. van Druten, D. Durfee, D. Kurn, and W. Ketterle,
Phys. Rev. Lett. \textbf{75}, 3969 (1995).
\bibitem{spinflip1997} C. V. Sukumar and D. M. Brink, Phys. Rev. A \textbf{56}, 2451 (1997).
\bibitem{kleppner1981} Jan R. Rubbmark, Michael M. Kash, Michael G. Littman, and Daniel Kleppner,
Phys. Rev. A \textbf{23}, 3107 (1981).
\bibitem{close2002}J. E. Lye, C. S. Fletcher, U. Kallmann, H-A Bachor and J. D. Close,
J. Opt. B \textbf{4}, 57 (2002).
\end{references}
\end{document}